\documentclass[10pt,conference]{IEEEtran}
\IEEEoverridecommandlockouts
\usepackage{cite}
\usepackage{braket}
\usepackage{amsmath,amssymb,amsfonts}
\usepackage{algorithmic}
\usepackage{graphicx}
\usepackage[hidelinks]{hyperref}
\usepackage{textcomp}
\usepackage{xcolor}
\def\BibTeX{{\rm B\kern-.05em{\sc i\kern-.025em b}\kern-.08em
T\kern-.1667em\lower.7ex\hbox{E}\kern-.125emX}}
\begin{document}

\title{Second-Order FALQON Parameter Transfer for the Max-Cut Problem on 3-Regular Graphs
  \thanks{This work was executed under the TIC26 -- Brazil Quantum Camp project, funded within the scope of the Prioritized Informatics Programs and Projects (PPI), Process No. 01245.008254/2025-22, under the responsibility of the Ministry of Science, Technology and Innovation (MCTI), with operational coordination by the Association for the Promotion of Brazilian Software Excellence (SOFTEX), and executed by CESAR and the Instituto de Pesquisas Eldorado.}
}

\author{
  \IEEEauthorblockN{Gabriel Fernandes Thomaz}
  \IEEEauthorblockA{\textit{Instituto de Pesquisas Eldorado}\\
    Porto Alegre, Brazil \\
  0009-0007-7410-5189}
  \and
  \IEEEauthorblockN{Eduarda Rodrigues Monteiro}
  \IEEEauthorblockA{\textit{PUCRS}\\
  Porto Alegre, Brazil}
  \IEEEauthorblockA{\textit{Instituto de Pesquisas Eldorado}\\
    Porto Alegre, Brazil \\
  0000-0003-2660-3910}
  \and
  \IEEEauthorblockN{Jerusa Marchi}
  \IEEEauthorblockA{\textit{UFSC}\\
  Florianópolis, Brazil}
  \IEEEauthorblockA{\textit{Instituto de Pesquisas Eldorado}\\
    Porto Alegre, Brazil \\
  0000-0002-4864-3764}
  \and
  \IEEEauthorblockN{Marcelo Zen Pretto}
  \IEEEauthorblockA{\textit{Instituto de Pesquisas Eldorado}\\
    Porto Alegre, Brazil \\
  0009-0004-9416-9794}
  \and
  \IEEEauthorblockN{Alisson dos Passos Fumaco}
  \IEEEauthorblockA{\textit{Instituto de Pesquisas Eldorado}\\
    Porto Alegre, Brazil \\
  0009-0003-3258-5146}
  \and
  \IEEEauthorblockN{Evandro Chagas Ribeiro da Rosa}
  \IEEEauthorblockA{\textit{UFSC}\\
  Florianópolis, Brazil}
  \IEEEauthorblockA{\textit{Instituto de Pesquisas Eldorado}\\
    Porto Alegre, Brazil \\
  0000-0002-8197-9454}
}

\maketitle

\begin{abstract}
  The Feedback-based Algorithm for Quantum Optimization (FALQON) offers a deterministic alternative to variational quantum algorithms by bypassing classical optimization loops. However, maintaining convergence on large problem instances often requires restricting the time step, necessitating quantum circuit depths that exceed Noisy Intermediate-Scale Quantum (NISQ) hardware capabilities. This paper investigates the parameter transferability of second-order FALQON applied to the Max-Cut problem on 3-regular graphs. Through numerical experiments evaluating quantum circuits up to 16 layers on graphs up to 24 nodes, we demonstrate a highly advantageous scaling behavior: transferring feedback parameters optimized on small instances to larger target graphs yields significantly higher approximation ratios than natively optimizing the parameters directly on the larger graphs. This performance advantage arises because parameters trained on smaller instances can safely adopt aggressively larger time steps. By offloading the expensive parameter discovery phase to small-scale instances, this transfer strategy simultaneously reduces computational overhead and enhances the approximation ratio, thereby bringing FALQON closer to practical viability on near-term quantum architectures.
\end{abstract}


\section{Introduction}
\label{sec:introduction}

Solving optimization problems more efficiently than classical counterparts remains a central objective of quantum computing. Variational quantum algorithms, such as the Quantum Approximate Optimization Algorithm (QAOA)~\cite{Farhi2014QAOA} and the Variational Quantum Eigensolver (VQE)~\cite{peruzzo_variational_2014}, tackle this using a hybrid classical-quantum approach. In this framework, the problem's cost function is encoded into a target Hamiltonian, and the optimization is performed by iteratively adjusting the parameters of a quantum circuit to minimize the expectation value of this Hamiltonian. However, the convergence and the quality of the final state are sensitive to the initial parameter guess and the efficiency of the classical optimizer navigating the resulting energy landscape. Alternatively, the Feedback-based Algorithm for Quantum Optimization (FALQON)~\cite{Magann2022FALQON} offers a deterministic approach that avoids classical optimization loops altogether by utilizing quantum feedback. This determinism, however, comes at a computational cost, as generating the feedback parameters requires evaluating the expectation values of Hamiltonians containing multiple non-commuting terms.

The evolution of the quantum state in FALQON is governed by a discrete time step, $\Delta t$. For an execution with $l$ layers, the algorithm must sequentially calculate $l$ feedback parameters, $\beta_k$, which requires evaluating the intermediate quantum state through circuits of progressively increasing depth. This parameter discovery phase constitutes the most computationally demanding step of the algorithm; computing these feedback parameters natively becomes prohibitively expensive for large problem instances, whether simulated classically or executed on a physical quantum processing unit (QPU).

Parameter transfer is a technique previously evaluated for QAOA to reduce training overhead by applying parameters optimized for one problem instance to another. If a sequence of parameters calculated for one instance can be effectively applied to another, the expensive feedback loop can be skipped for the target problem. This advantage is maximized if the optimal parameters can be discovered on a small, computationally inexpensive instance and subsequently transferred to a larger, more complex one.

In this paper, we evaluate this parameter transfer technique for the second-order FALQON~\cite{arai_scalable_2025} applied to the Max-Cut problem. Our findings demonstrate that transferring the $\Delta t$ and $\beta_k$ parameters from a smaller problem instance to a larger one not only eliminates the high cost of native parameter calculation but also yields significantly better approximation ratios than attempting to optimize the parameters directly on the larger instance. Specifically, we evaluate the transferability of second-order FALQON parameters for solving the Max-Cut problem on 3-regular graphs of varying node sizes.

The remainder of this paper is organized as follows: Section \ref{sec:falqon} presents the second-order FALQON algorithm applied to the Max-Cut problem. Section \ref{sec:related_works} reviews related work on parameter transfer in quantum algorithms. Section \ref{sec:experiments} details the methodology for the numerical experiments, and Section \ref{sec:analysis} provides an analysis of the results. Finally, Section \ref{sec:conclusion} offers concluding remarks.

\section{Second-Order FALQON for Max-Cut}
\label{sec:falqon}

This section introduces the second-order Feedback-based Algorithm for Quantum Optimization (FALQON)~\cite{Magann2022FALQON} in the context of the Max-Cut problem. An approach that incorporates higher-order corrections to improve stability and reduce circuit depth, making it particularly suitable for instances of large-scale problems and subsequent parameter transfer strategies.

\subsection{Max-Cut as a Quantum Optimization Problem}

The Max-Cut problem is defined on an undirected graph $G = (V,\ E)$, where the objective is to partition the set of vertices into two disjoint subsets such that the number of edges crossing the partition is maximized. This problem is NP-hard and has become a standard benchmark for quantum optimization algorithms. In the quantum setting, Max-Cut is canonically formulated through a cost Hamiltonian of the form
\begin{equation}
  H_C = -\frac{1}{2}\sum_{(i,\ j)\in E} \left(1 - Z_i Z_j\right),
\end{equation}
where $Z_i$ denotes the Pauli-$Z$ operator acting on qubit $i$~\cite{Farhi2014QAOA}. The ground state of this Hamiltonian encodes the optimal solution to the Max-Cut problem.

While variational algorithms like QAOA are commonly used to tackle this Hamiltonian, their reliance on classical optimization loops presents scalability challenges~\cite{McClean2018BarrenPlateaus}, motivating the use of deterministic, feedback-driven alternatives.

\subsection{Feedback-Based Quantum Optimization}

An alternative to variational optimization is the Feedback-based Algorithm for Quantum Optimization (FALQON). Instead of employing a classical optimizer to search an energy landscape, FALQON deterministically constructs a quantum circuit by iteratively updating feedback parameters based on measurement outcomes obtained from the intermediate quantum state~\cite{Magann2022FALQON}. The circuit is generated through successive applications of unitary evolutions governed by a combination of a mixing Hamiltonian $H_M = \sum X_i$ and the problem Hamiltonian $H_C$.

The state evolution at the $k$-th layer can be written as
\begin{equation}
  \ket{\psi_{k+1}} = e^{-i\Delta t \beta_k H_M} e^{-i\Delta t H_C} \ket{\psi_{k}},
\end{equation}
where $\Delta t$ is a fixed time step and $\beta_k$ is a feedback parameter determined from measurements on the current state $\ket{\psi_k}$.

The feedback rule is derived from a Lyapunov-control-inspired framework, setting
\begin{equation}
  \beta_k = -\langle A \rangle_k,
\end{equation}
where $\langle A \rangle_k = \bra{\psi_k} A \ket{\psi_k}$ is the expectation value of the commutator operator
\begin{equation}
  A = i[H_M, H_C].
\end{equation}
This assignment guarantees a monotonic improvement of the cost function in the limit of a sufficiently small $\Delta t$~\cite{Magann2022FALQON}.

This deterministic construction eliminates the need for repeated classical optimization loops, distinguishing FALQON from QAOA while retaining a layered circuit structure compatible with near-term quantum hardware~\cite{Farhi2017FixedArch}.

The original FALQON formulation can be interpreted as a first-order discretization of a continuous-time control process. As such, its convergence guarantees rely heavily on the assumption of a small time step $\Delta t$. While small values of $\Delta t$ ensure stability and a strict monotonic decrease of $\langle H_C \rangle$, they necessitate a prohibitively large number of layers to reach states with high approximation ratios. Conversely, increasing $\Delta t$ accelerates convergence but may lead to oscillatory or unstable behavior due to neglected higher-order terms in the time-discretized evolution. This trade-off between stability and circuit depth constitutes a primary practical limitation of first-order FALQON implementations, particularly for larger problem instances~\cite{Magann2021PRXQuantum}.

\subsection{Second-Order FALQON}

Second-order FALQON~\cite{arai_scalable_2025} addresses the limitations of the first-order approach by incorporating higher-order contributions into the discretized time evolution. Rather than relying solely on a linear expansion in $\Delta t$, the second-order formulation accounts for quadratic terms associated with the curvature of the energy landscape. From a numerical integration perspective, this corresponds to a higher-order approximation of the underlying continuous dynamics, which mitigates discretization errors and enables stable state evolution over larger time steps.

In practical terms, second-order FALQON modifies the feedback law by extracting additional information from nested commutators involving the cost Hamiltonian ($H_C$) and the mixing Hamiltonian ($H_M$). This captures both the gradient and the curvature of the objective function. The updated feedback parameter $\beta_k$ for a given layer is defined as:
\begin{equation}
  \beta_k = -\frac{\langle A \rangle_k + \Delta t \langle C \rangle_k}{2\Delta t \langle B \rangle_k},
\end{equation}
where $A$ represents the first-order gradient operator, and the higher-order operators are given by:
\begin{align}
  B & = \frac{1}{2}[[H_M, H_C], H_M],\\
  C & = [[H_M, H_C], H_C].
\end{align}

By utilizing this enhanced feedback mechanism, the algorithm can take larger effective steps along the optimization trajectory without sacrificing its stability or monotonic convergence. When applied to the Max-Cut problem, this translates to a significant reduction in the required circuit depth. Furthermore, the ability to operate stably with a larger $\Delta t$ is precisely what makes second-order FALQON exceptionally well-suited for parameter transfer strategies. It allows the aggressive time steps discovered on computationally tractable small graphs to successfully drive the evolution of large-scale instances, where the cost of native parameter discovery would otherwise be prohibitive.

\section{Related Works}
\label{sec:related_works}

Hybrid quantum algorithms, particularly variational quantum algorithms (VQAs), have been widely investigated as candidates for near-term combinatorial optimization. The Quantum Approximate Optimization Algorithm (QAOA) is a representative example, alternating between problem and mixer Hamiltonians in a parameterized circuit optimized through a classical outer loop. Despite promising results, the scalability of VQAs is often limited by difficult optimization landscapes and sensitivity to noise, especially at increasing circuit depths.

To circumvent these limitations, feedback-based approaches inspired by quantum control have been proposed. The Feedback-based Algorithm for Quantum Optimization (FALQON) replaces classical optimization with deterministic, measurement-informed parameter updates derived from commutators between the cost and driver Hamiltonians, guaranteeing monotonic improvement under suitable conditions. While this approach eliminates the classical optimization loop, the computational cost of discovering feedback parameters remains a bottleneck for large problem instances.

In parallel, a substantial body of work has explored \emph{parameter transfer} in variational settings, particularly for QAOA, showing that optimized parameters often exhibit transferable structure across related problem instances \cite{galda2023similarity}. Such transfer strategies reduce training overhead and improve robustness, motivating their adoption in scalable quantum optimization.

Recent work has further strengthened the role of parameter transfer as a practical tool for scaling variational quantum algorithms. In particular, Kotil \emph{et al.} demonstrated that QAOA parameters trained on small problem instances can be successfully transferred to larger instances in the context of multi-objective weighted Max-Cut, substantially reducing the need for retraining on large quantum systems~\cite{Kotil2025MultiObjectiveQAOA}. By leveraging parameter reuse, their approach significantly alleviates training overhead while preserving solution quality, highlighting parameter transfer as a key mechanism to overcome scalability bottlenecks in near-term quantum optimization.

In contrast, parameter transfer in feedback-based quantum optimization has received comparatively little attention, with existing approaches largely relying on learning-based techniques to predict feedback schedules \cite{perez2026learning}. This work addresses this gap by investigating direct parameter transfer in second-order FALQON, enabling efficient reuse of feedback parameters across Max-Cut instances without classical re-optimization or learning-based models.

\section{Numerical Experiments}
\label{sec:experiments}

To evaluate the parameter transferability of the FALQON algorithm, we generated a dataset of random 3-regular graphs with node sizes ranging from $n=6$ to $n=24$. For each specific node size, an ensemble of 20 distinct graphs was generated. The exact ground state energy for the Max-Cut problem was computed classically for each instance using simulated annealing to establish a baseline, enabling the calculation of the exact approximation ratio. For the subset of graphs with node sizes $n_{\text{train}} \in \{6, 8, \dots, 18\}$, we optimized the feedback parameters and subsequently performed a cross-evaluation by transferring these parameters to target graphs of size $n_{\text{target}}$, strictly where $n_{\text{train}} \le n_{\text{target}}$.

\begin{figure}[htbp]
  \centering
  \includegraphics[width=\linewidth]{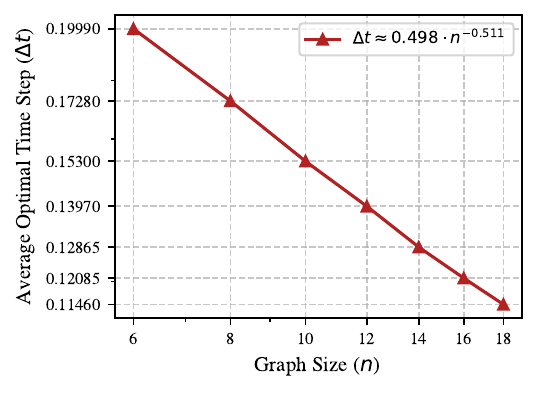}
  \caption{Average optimal time step ($\Delta t$) that maximizes the approximation ratio after 16 layers, plotted on a log-log scale as a function of the 3-regular graph size ($n$). The monotonically decreasing trend exhibits a strict power-law decay with a fitted regression of $\Delta t \approx 0.4984 \cdot n^{-0.5110}$, indicating that the time step scales approximately as $1/(2\sqrt{n})$.}  \label{fig:deltat}
\end{figure}

The parameter optimization phase consisted of simulating the FALQON algorithm for a fixed depth of $l = 16$ layers. For each training graph, we performed a linear scan of the time step, $\Delta t$, over the interval $[0.1, 1.0]$ with a resolution of $0.001$. From this parameter sweep, we extracted the specific $\Delta t$ and its corresponding generated sequence of feedback parameters, $\beta_k$, that maximized the approximation ratio at the final layer. To reduce the computational overhead of the scan, an early stopping criterion was implemented: the search was terminated once the incremented $\Delta t$ value caused the algorithm to diverge or fail to meet a minimum convergence threshold. The optimal $\Delta t$ values, averaged across the 20 instances for each respective node size, are presented in Figure \ref{fig:deltat}.

A log-log regression of these empirical results demonstrates a strict power-law scaling governed by the relationship
\begin{equation}
  \Delta t \approx 0.4984 \cdot n^{-0.5110}.
\end{equation}
This closely approximates a scaling of $\Delta t \approx \frac{1}{2\sqrt{n}}$, indicating that the required temporal resolution tightens inversely with the square root of the graph size. This empirical power-law highlights the exact nature of the growing bottleneck: as the system size increases, the native parameter discovery is forced to take increasingly smaller steps, severely limiting exploration within a fixed circuit depth.

After completing the native parameter optimization, we evaluated the transferability of the results. The optimal sequence $(\Delta t,\ \beta_k)$ obtained from a specific training graph was applied directly to the problem Hamiltonian of a target graph. We quantified the success of this transfer by calculating the approximation ratio, defined as the final quantum expectation value produced by the transferred parameters divided by the target graph's exact ground state energy. For each valid size pair $(n_{\text{train}}, n_{\text{target}})$, this cross-evaluation was performed across all 400 combinations of the 20 training graphs and 20 target graphs.

\begin{figure}[htbp]
  \centering
  \includegraphics[width=\linewidth]{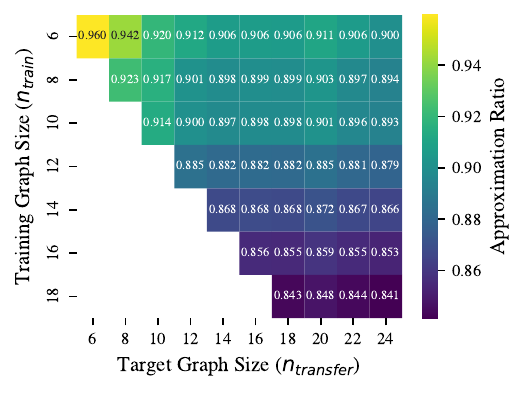}
  \caption{FALQON parameter transferability matrix. The heatmap displays the average approximation ratio achieved when evaluating 16-layer FALQON circuits using parameters optimized on training graphs ($n_{\text{train}}$, y-axis) applied to the Hamiltonians of target graphs ($n_{\text{target}}$, x-axis).}
  \label{fig:heatmap}
\end{figure}

The overall results were aggregated by averaging the approximation ratios for each size pair. The complete transfer matrix, detailing the average approximation ratio achieved when mapping parameters from $n_{\text{train}}$ to $n_{\text{target}}$, is illustrated as a heatmap in Figure \ref{fig:heatmap}.

\vspace{1cm}
\section{Result Analysis}
\label{sec:analysis}

\begin{figure*}
  \includegraphics[width=\linewidth]{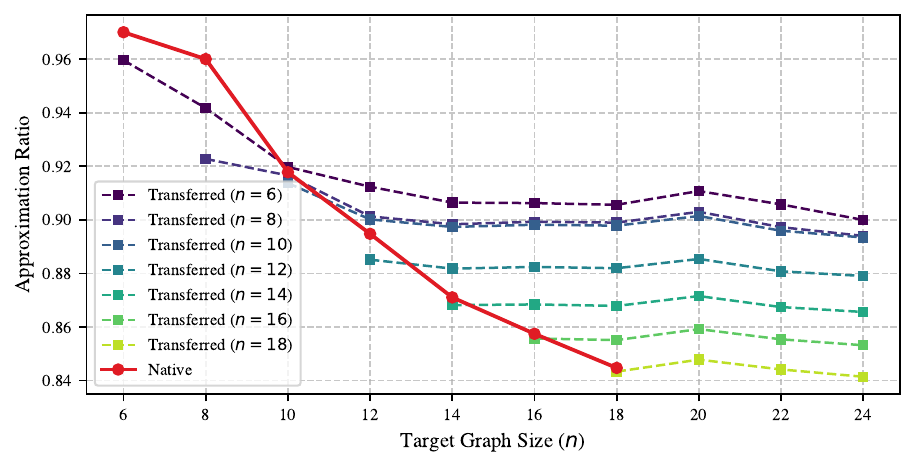}
  \caption{Approximation ratio of natively optimized parameters (solid red line) versus parameters transferred from smaller instances (dashed lines). The native performance decays significantly as the target graph size ($n$) increases due to the algorithm selecting progressively smaller time steps ($\Delta t$) that restrict exploration within the fixed 16-layer limit. In contrast, transferring parameters from smaller graphs (particularly $n = 6$) significantly outperforms native optimization on large graphs.}
  \label{fig:performance}
\end{figure*}

The experimental results reveal a non-intuitive scaling behavior regarding the FALQON algorithm's parameter transferability. As indicated by the matrix in Figure~\ref{fig:heatmap}, transferring parameters optimized on a smaller source graph to a larger target graph often yields superior performance compared to optimizing the feedback parameters natively on the target graph itself. This phenomenon is further detailed in Figure~\ref{fig:performance}, where the solid red line represents the baseline approximation ratio achieved through native optimization. In contrast, the dashed lines illustrate the approximation ratios achieved when parameters from smaller training graphs are applied to larger targets. Notably, the parameters trained on $n=6$ (the smallest instances evaluated) consistently produced the highest approximation ratios across all evaluated target sizes.

Our initial hypothesis was that transferred parameters would provide a ``good enough'' baseline approximation ratio that would steadily degrade as the structural disparity (node count) between the source and target graphs increased. However, the numerical results contradict this expectation; the transferred approximation ratios remain remarkably stable across target sizes, exhibiting only a marginal decay.

We hypothesize that this transferability advantage is driven by the dynamic selection of the time step, $\Delta t$. As the graph size increases, the overall energy scale (or spectral norm) of the problem Hamiltonian also increases. To ensure convergence and prevent overshooting the target state trajectory, the native optimization procedure must select progressively smaller values for $\Delta t$. This decrease is clearly visualized in Figure~\ref{fig:deltat}.

While a smaller $\Delta t$ guarantees theoretical convergence in the infinite-depth limit, it becomes a severe bottleneck within a constrained circuit depth. Because our experiments were strictly limited to $l=16$ layers, the algorithm operating with natively optimized, small time steps fails to reach high-quality solutions, resulting in the steady decline of the native approximation ratio (solid red line, Figure~\ref{fig:performance}).

Conversely, when parameters are transferred from a smaller graph (\textit{e.g.}, $n=6$), the applied $\Delta t$ is significantly larger (approximately $0.20$ versus $0.115$ for larger graphs). These aggressive, large-step-size parameters, which the optimization routine ``safely'' learned on the smaller energy landscape, act as an accelerated evolutionary path. When applied to larger graphs, this sequence forces the state to evolve more rapidly per layer, achieving much higher approximation ratios within the strict 16-layer limit.

Historically, the transition from first-order to second-order discretization in FALQON was motivated precisely by the ability to utilize larger $\Delta t$ steps, albeit at the cost of evaluating higher-order nested commutators. The results presented here suggest a different paradigm for algorithmic improvement. By exploiting parameter transferability, we achieve the benefits of an aggressively large $\Delta t$ while simultaneously \textit{reducing} the computational overhead, as the expensive feedback parameter discovery is relegated to a small instance.

\section{Final Remarks}
\label{sec:conclusion}

This paper presented numerical results indicating that parameter transfer offers a highly effective pathway to enhance the scalability of the FALQON algorithm. Although our numerical experiments were bounded to target graphs of up to 24 nodes, the data reveals a compelling trend: training the FALQON algorithm on a small problem instance and transferring the resulting feedback parameters to a larger target instance yields significantly higher approximation ratios than calculating those parameters natively on the larger instance.

This transferability advantage was specifically evaluated for the Max-Cut problem on topologies that share a uniform local structure, namely 3-regular graphs. It is important to note that this success relies on local structural similarities and will likely not generalize as effectively to topologies with highly variable degree distributions, such as sparse Erdős-Rényi graphs.

Most impactfully, these findings suggest a paradigm shift in how FALQON feedback parameters are generated. By enabling the successful application of larger time steps ($\Delta t$) learned from smaller graphs, parameter transfer accelerates state convergence, which in turn reduces the required circuit depth (number of layers). Simultaneously, it reduces the computational overhead of the parameter discovery phase, bringing FALQON closer to the stringent depth and coherence constraints of Noisy Intermediate-Scale Quantum (NISQ) devices. Furthermore, confining the parameter calculation to small problem sizes allows this phase to be entirely offloaded to classical simulators, thereby isolating the parameter generation from the intrinsic noise of physical QPU executions.

As this study presents initial evidence, it opens avenues for future investigation:
\begin{itemize}
  \item \emph{Theoretical Foundations:} Develop a rigorous analytical framework to explain the underlying mechanics of parameter transfer to formally prove how local graph symmetries govern the global optimization trajectory.
  \item \emph{Scalability and Generality:} Expand the numerical evaluations to significantly larger graphs, and assess parameter transferability across different graph classes and combinatorial optimization problems beyond Max-Cut.
  \item \emph{Noise Resilience:} Evaluate the impact of quantum hardware noise on both the parameter training process and the execution of the transferred parameter sequences.
  \item \emph{Algorithmic Graph Reduction:} Given a large, arbitrary problem instance, explore systematic methodologies to derive a small representative training graph that maximizes parameter transfer success.
\end{itemize}

\subsection*{Research Artifacts}

To facilitate reproducibility, all code and datasets generated during this study have been made available at \url{https://github.com/eldorado-institute}.

% https://github.com/Brazil-Quantum-Camp/FALQON-Parameter-Transfer
% \section*{Acknowledgment}
\bibliographystyle{IEEEtran}
\bibliography{main_bibliography}

\end{document}